\documentclass{emulateapj}
\usepackage{natbib}
\usepackage{epsfig}

\shorttitle{Radio Spectrum of U/LIRGS}
\shortauthors{Leroy et al.}

\begin{document}

\slugcomment{Draft} \title{Complex Radio Spectral Energy Distributions in Luminous and Ultraluminous Infrared Galaxies}

\author{
Adam K. Leroy\altaffilmark{1,8}, 
Aaron S. Evans\altaffilmark{1,3},
Emmanuel Momjian\altaffilmark{2}, 
Eric Murphy\altaffilmark{4}, 
J\"urgen Ott\altaffilmark{2}, 
Lee Armus\altaffilmark{5}, 
James Condon\altaffilmark{1},
Sebastian Haan\altaffilmark{5}, 
Joseph M. Mazzarella\altaffilmark{5},
David S. Meier\altaffilmark{6,2}, 
George C. Privon\altaffilmark{3}, 
Eva Schinnerer\altaffilmark{7}, 
Jason Surace\altaffilmark{5}, 
Fabian Walter\altaffilmark{7}}

\altaffiltext{1}{National Radio Astronomy Observatory, 520 Edgemont Road, Charlottesville, VA, 22903-2475, USA}
\altaffiltext{2}{National Radio Astronomy Observatory, P. O. Box O, Socorro, NM 87801, USA}
\altaffiltext{3}{Department of Astronomy, University of Virginia, 530 McCormick Road, Charlottesville, VA 22904}
\altaffiltext{4}{Observatories of the Carnegie Institution for Science, 813 Santa Barbara Street, Pasadena, CA 91101, USA}
\altaffiltext{5}{Spitzer Science Center, California Institute of Technology, MC 314-6, Pasadena, CA 91125, USA}
\altaffiltext{6}{New Mexico Institute of Mining and Technology, 801 Leroy Place, Socorro, NM 87801}
\altaffiltext{7}{Max Planck Institut für Astronomie, Königstuhl 17, Heidelberg D-69117, Germany}
\altaffiltext{8}{Hubble Fellow}


\begin{abstract}
We use the Expanded Very Large Array to image radio continuum emission
from local luminous and ultraluminous infrared galaxies (LIRGs and
ULIRGs) in 1 GHz windows centered at 4.7, 7.2, 29, and 36~GHz.  This
allows us to probe the integrated radio spectral energy distribution
(SED) of the most energetic galaxies in the local universe. The
4--8~GHz flux densities agree well with previous measurements. They
yield spectral indices $\alpha \approx -0.67$ (where $F_{\rm \nu}
\propto \nu^{\alpha}$) with $\pm 0.15$ (1$\sigma$) scatter, typical of
nonthermal (synchrotron) emission from star-forming galaxies. The
contrast of our 4--8~GHz data with literature $1.5$ and $8.4$~GHz flux
densities gives further evidence for curvature of the radio SED of
U/LIRGs. The SED appears flatter near $\sim 1$~GHz than near $\sim
6$~GHz, suggesting significant optical depth effects at the lower
frequencies. The high frequency (28--37~GHz) flux densities are low
compared to extrapolations from the 4--8~GHz data. We confirm and
extend to higher frequency a previously observed deficit of high
frequency radio emission for luminous starburst galaxies.
\end{abstract}


\section{Introduction}
\label{sec:intro}

Most very luminous galaxies at $z \sim 0$ emit the bulk of their
luminosity in the infrared (IR). These luminous and ultraluminous
infrared galaxies (LIRGs and ULIRGs, infrared luminosity $L_{\rm IR}
\left[ 8-1000\mu {\rm m} \right] > 10^{11}$ and $L_{\rm IR} >
10^{12}$~$L_{\odot}$) make up the majority of galaxies with $L_{\rm
  bol} \gtrsim 10^{11.5}$~L$_\odot$.  These systems often correspond
to major mergers and compared to normal star-forming disk galaxies,
U/LIRGs exhibit enhanced, concentrated molecular gas content, high
star formation efficiency, increased active galactic nuclei (AGN)
activity, and disturbed morphologies \citep{SANDERS96}. Even the
closest ULIRGs are $\gtrsim 100$~Mpc away with compact radio, IR, and
millimeter line emission
\citep[e.g.,][]{LONSDALE93,DOWNES98,SAKAMOTO08}, so that high angular
resolution (such as that obtained by the EVLA) is required to study
their structure.

Nonthermal (synchrotron) radio emission is well-established to track
infrared emission \citep[the "radio-FIR
  correlation,"][]{HELOU85,CONDON92,YUN01} while thermal
(bremsstrahlung) radio emission at $\nu \gtrsim 4$~GHz gives an
unobscured view of ionizing photon production. These emission
mechanisms, the indifference of radio waves to dust, and the resolving
power of cm-wave interferometers make the radio continuum a natural
tool to dissect the detailed structure of U/LIRGs. Deviations from the
radio-IR correlation, radio morphology, and brightness can all help
distinguish the presence and contribution of AGN to the overall
luminosity, though not necessarily uniquely \citep[][and references in
  the latter]{CONDON91,LONSDALE95,SAJINA08,LONSDALE06}.

This letter reports first results from our Expanded Very Large Array
(EVLA) program to explore the structure of star formation in U/LIRGs
using radio continuum emission. We are imaging 22 of the most luminous
nearby U/LIRGs ($L_{\rm IR} > 10^{11.6}$~L$_\odot$) in the C (4-8~GHz)
band and Ka (26.5-40~GHz) bands. Here we report observations from the
EVLA's C and D configurations. The EVLA's continuum sensitivity makes
it possible to achieve good signal-to-noise ratio, even at high
frequencies, with modest time on source. The large instantaneous
bandwidth allows us to probe the radio spectral index, $\alpha$
(defined so that $F_{\rm \nu} \propto \nu^{\alpha}$), both within and
between bands.

Here explore variations in the radio spectral energy distribution
(SED) of our targets and in the radio-IR correlation. In a simple
picture dominated by star formation we would expect the C band to be
dominated by synchrotron emission with spectral index $\alpha \approx
-0.8$ while the Ka contains an approximately equal mixture of
nonthermal emission and thermal emission, which has $\nu = -0.1$. For
systems dominated by an AGN the spectral index may show wider
variation depending on the orientation of the AGN or injection
spectrum of electron energies. We close by showing overlays of the
Ka-band emission on HST imaging. These highlight the EVLA's improved
ability to trace embedded star formation at high resolution.

\section{Observations and Data Reduction}
\label{sec:data}

Our EVLA resident shared risk observing program, AL746, is imaging
radio continuum emission from nearby LIRGs and ULIRGs. The sample,
listed in Table \ref{tab:results}, consists of 22 of the most luminous
members of the IRAS Revised Bright Galaxy Sample
\citep[RBGS][]{SANDERS03} northern enough to be observed by the
EVLA. Distances span $50$--$300$~Mpc, with a median of $177$~Mpc, and
IR luminosities go from $\log_{\rm 10} L_{\rm IR} \left[ {\rm
    L}_{\odot} \right] = 11.6$--$12.6$\footnote{This refers to
  estimates of the ``total'' (8-1000$\mu$m) infrared luminosity
  constructed from IRAS flux densities by \citet{SANDERS03} following
  \citet{SANDERS96}.}.

The whole program consists of imaging each target at four frequencies,
$\nu \sim 4.7$, $7.2$, $29$, and $36$~GHz, with a bandwidth of 1~GHz
centered at each frequency. We observe the two low frequency windows
(together the ``C-band'' data) and two high frequencies (the
``Ka-band'' data) simultaneously. Here we report the results from the
two most compact EVLA configurations (C and D). We observed each
target for $\sim 5$~minutes in each band and configuration.

We reduced the data using the Common Astronomy Software Applications
(CASA) package developed by NRAO. We adopted standard approaches to
flag bad data (radio frequency interference is a particular issue in
the C band) and calibrate the bandpass, phase, and amplitude response
of individual antennas. We iteratively self-calibrated the data
whenever possible.

We combined all data for each band into a single image using CASA's
multi-frequency synthesis (MFS) capabilities
\citep[see][]{SAULT94}. This image reconstruction mode synthesizes
interferometer observations across a wide bandwidth. It accounts for
changes in the synthesized beam as a function of frequency and
produces images of intensity for a fiducial frequency and maps of the
spectral index. For our data, these fiducial frequencies are 5.95~GHz
for the C band and 32.5~GHz for the Ka band. Note that we do not
actually observe at either of these frequencies, they are merely a
intermediate between our two separated 1~GHz windows. At C-band, we
deconvolved (CLEAN) the image using components that have both a flux
density and a spectral index. At Ka band we use components that have
only flux density at the fiducial frequency because we expect a
flatter spectral index, have a lower signal-to-noise ratio, and have a
smaller fractional bandwidth.

After reduction, we have maps of intensity for C and Ka band and
spectral index at C band. Synthesized beams are $\sim 8\arcsec$ at C
band and $\sim 1\arcsec$ at Ka band. The noise in a C-band image is
usually $\approx 20~\mu$Jy~beam$^{-1}$, though for our brightest
targets calibration issues sometimes limit us to about twice this
value. A typical noise in a Ka-band image is $\approx
50~\mu$Jy~beam$^{-1}$. All targets are detected, with peak SNR from
$200$ to $6.3 \times 10^3$ at C band (median 670) and 12 to 500 at Ka
band (median 50).

We measure C-band flux densities in the following way: we estimate the
RMS noise away from the source. We then identify the region of
contiguous SNR $>10$ emission that contains the galaxy. We integrate
over this area to derive the flux density. We take the
intensity-weighted mean of the spectral index $\left< \alpha_{\rm C}
\right> = \int F_{\nu} \alpha / \int F_{\nu}$ derived by the MFS CLEAN
as the spectral index for the source. We derive Ka-band flux densities
by integrating over the same area used to derive the C-band flux
densities. Over this matched area, two galaxies have integrated SNR $<
3$ at Ka and we omit them from all Ka-band related analysis.

\section{Results}
\label{sec:results}

\begin{figure*}
\plottwo{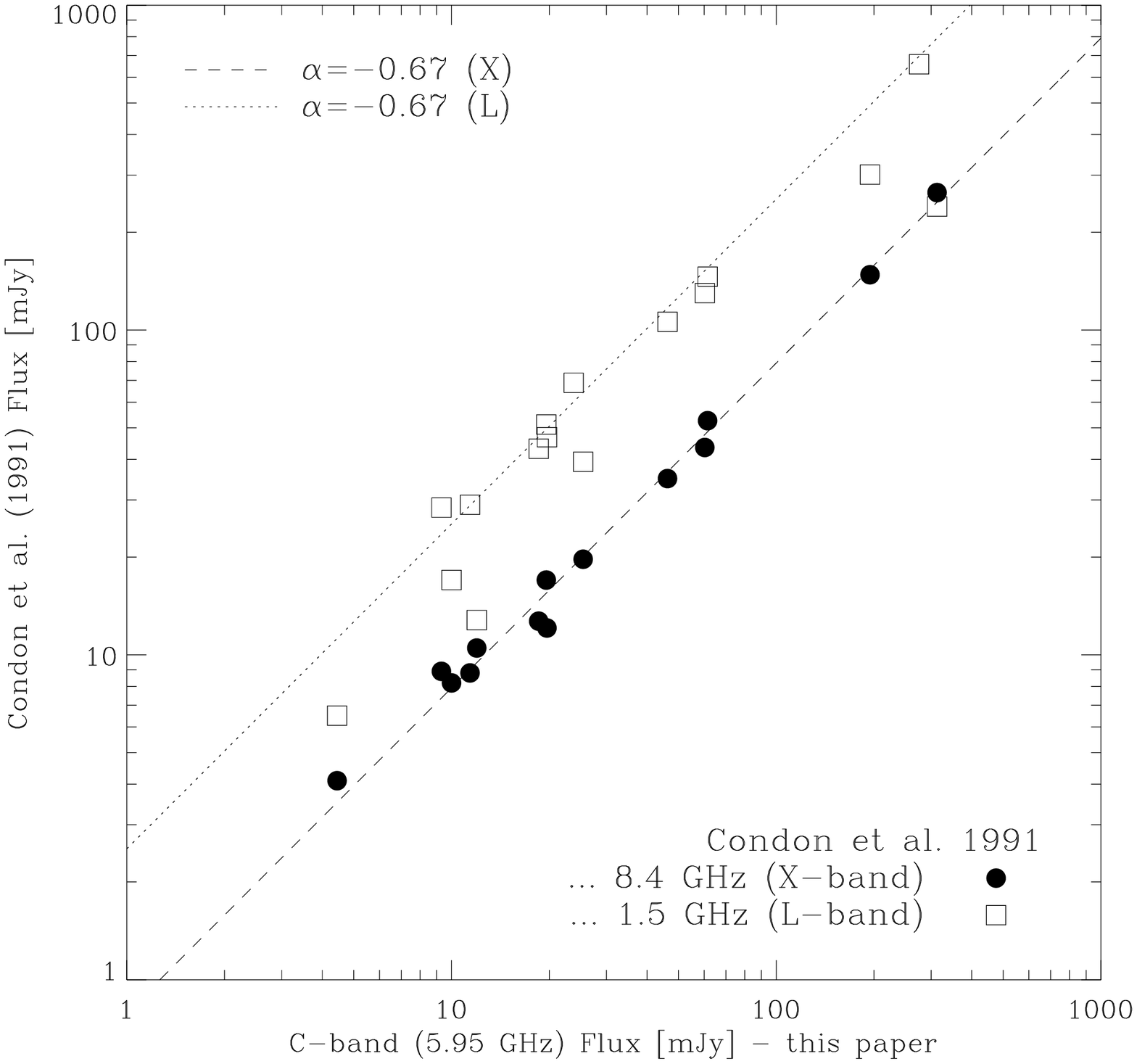}{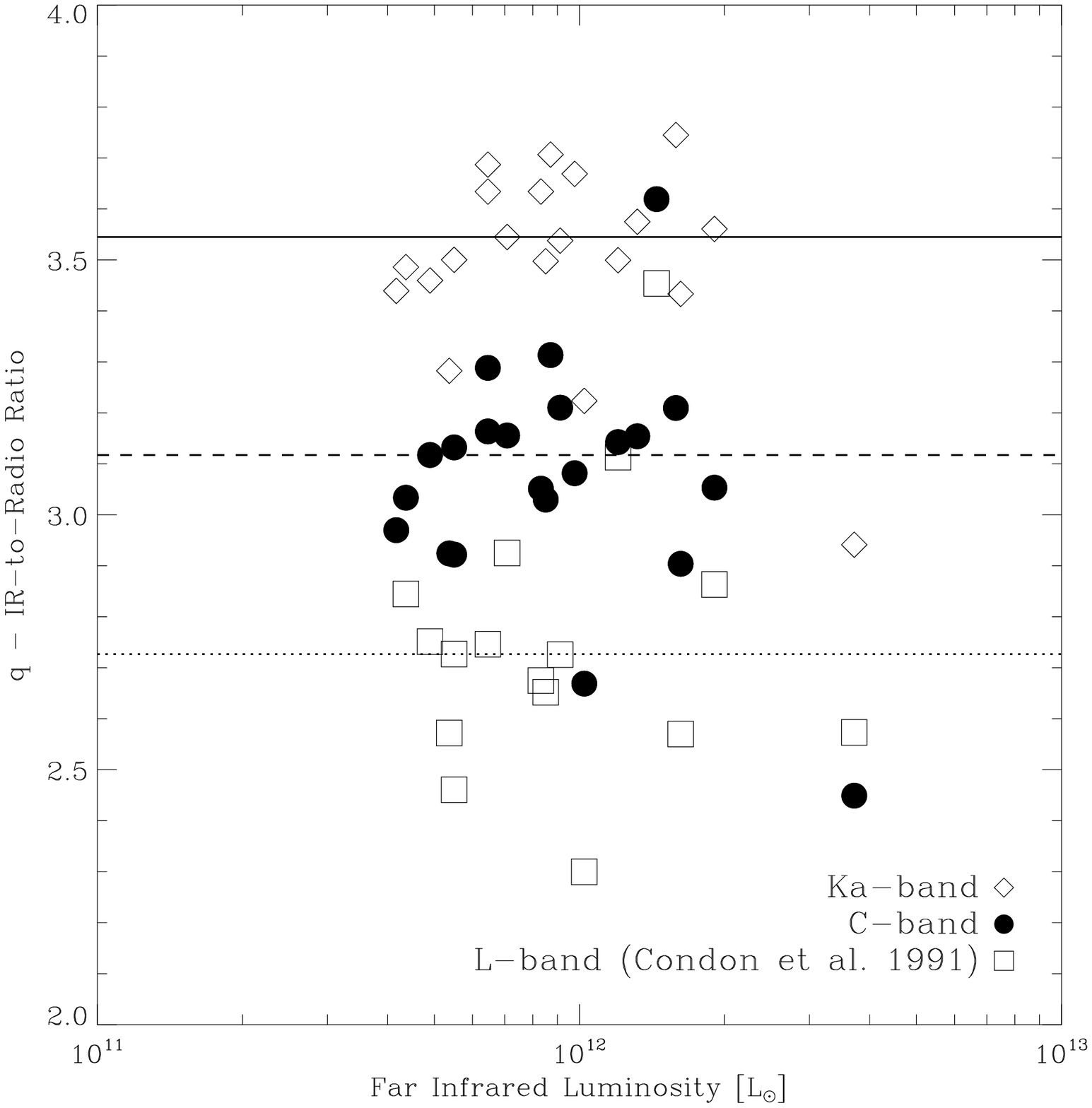}
\caption{\label{fig:cband} ({\em left}) Flux densities from
  \citet[][]{CONDON91} ($y$-axis) as a function of our 5.95~GHz
  (C-band) flux density ($x$-axis). Filled circles show X-band (8.4
  GHz) flux densities. Open squares show L-band (1.5 GHz) flux
  densities. Dashed lines show the expected flux density given our
  C-band flux density and a spectral index $\alpha = -0.67$. Our
  C-band flux densities agree well with the X-band values. The L-band
  flux densities tend to lie below the expected value, indicating
  curvature in the radio SED in the range $1$--$5$~GHz.  ({\em right})
  Ratio of far-infrared to radio emission ($q$) for Ka (open
  diamonds), C (black circles), and L (open squares) bands. All three
  bands correlate with IR emission, as expected, with the lowest
  scatter in the C-band. The highest luminosity source, with an
  inverted spectrum, is UGC 08058 (Markarian 231), which is dominated
  by an AGN. Red outlines highlight source and two other outliers
  discussed in the text.}
\end{figure*}

\subsection{A Typical Synchrotron Spectrum at C-Band}

The left panel in Figure \ref{fig:cband} compares our measured C-band
flux densities at 5.95 GHz ($x$-axis) to previous measurements at
8.4~GHz (black circles) and 1.5~GHz (open squares) both drawn from
\citet{CONDON91}. Table \ref{tab:results} reports internal C-band
spectral indices, $\alpha_{\rm C}$. Our median $\alpha_{\rm C}$ is
$-0.67$ with $1\sigma$ scatter $\approx 0.15$. The SNR of the data are
very high but an uncertainty of even a few percent in the relative
calibrations of the two parts of the C-band (e.g., due to calibration
problems, radio frequency interference, uncertainties in the primary
calibrator model, or CLEAN issues) will lead to uncertainties $\sim
0.1$.

We expect nonthermal (synchrotron) emission to dominate the
C-band. Our measured $\alpha = -0.67$ is close to a normal synchrotron
spectral index for a star-forming galaxy, $\alpha \approx -0.8$
\citep{CONDON92} though AGN can produce similar indices. For
$\alpha_{\rm C} = -0.67$, Figure \ref{tab:results} shows that our
C-band flux densities agree with the X-band measurements of
\citet{CONDON91} well. The median ratio after correcting to the same
band using $\alpha_{\rm C} = -0.67$ is $\approx 1.02$ and the scatter
is $14\%$, indicating very good agreement. Thus our C-band
measurements from $\approx 4.2$--$5.2$ and $6.7$--$7.7$~GHz and
previous measurements at $8.44$~GHz all exhibit good consistency with
a typical synchrotron spectrum.

\begin{figure*}
\plotone{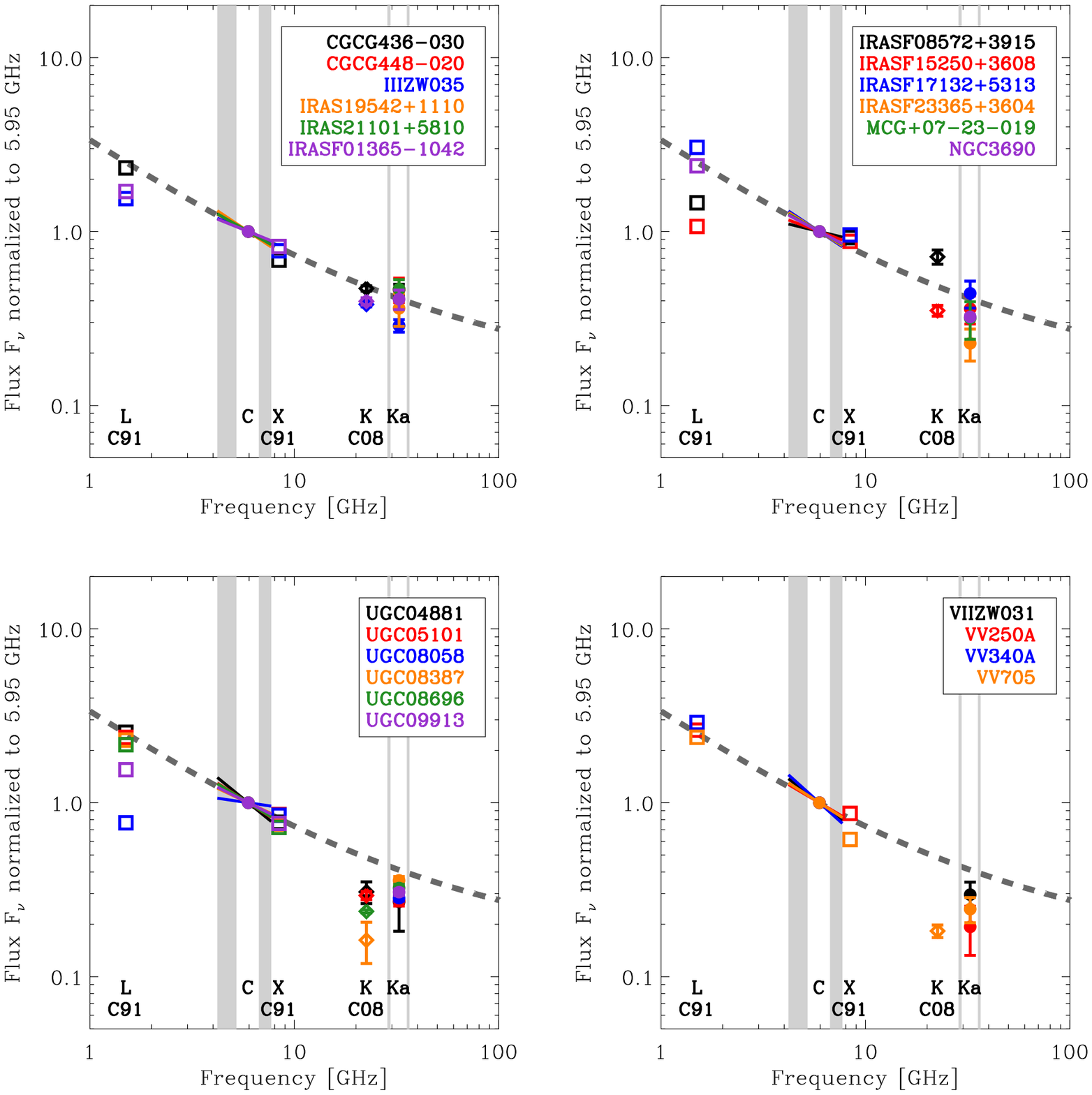}
\caption{\label{fig:seds} Normalized spectral energy distributions of
  our targets, sorted by IR luminosity from top left to bottom
  right. Gray regions show our observed frequencies and circles
indicate our measured flux densities. Lines show $\alpha_{\rm C}$
across our C-band measurement. We normalize all values to the 5.95~GHz
value and plot literature measurements as open symbols. The dashed
line shows a simple two-component model mixing optically thin
synchrotron ($\alpha = -0.8$) with thermal emission ($\alpha = -0.1$)
at levels appropriate for a star-forming galaxy \citep{CONDON92}. Our
C-band measurements show an approximately normal synchrotron shape in
most sources, but U/LIRGs depart from this simple model at low and
high frequencies. Our new Ka measurements confirm a systematic deficit
of high-frequency emission, extending this result to higher
frequencies ($\nu \sim 36$~GHz).}
\end{figure*}

\subsection{Departures From a Simple SED at Low and High $\nu$} 

In the simple two-component model outlined in the introduction
synchrotron emission with $\alpha \sim -0.8$ dominates below $\nu \sim
10$~GHz. Above this frequency, the radio continuum results from a
mixture of synchrotron and thermal emission, which has $\alpha =
-0.1$. In a normal star-forming galaxy the ratio of these two
processes may be approximately fixed. Figure \ref{fig:seds} shows the
radio SEDs of our targets, which exhibit significant departures from
this picture at both high and low frequency. We normalize each SED at
5.95~GHz, show literature flux densities from \citet{CONDON91} and
\citet{CLEMENS08}, and overplot the simple two-component, optically
thin model from \citet{CONDON92}. The lines through our C-band points
show our measured $\alpha_{\rm C}$.

At low frequencies, the 1.5~GHz (L band) flux densities lie mostly
below an extrapolation of the synchrotron emission measured in C-band,
which appears as a dashed line in Figure \ref{fig:seds}. This can be
seen from the left panel in Figure \ref{fig:cband} where the 1.5 GHz
points (open squares) tend to scatter below the value expected for our
measured $\alpha_{\rm C}$ (dashed). Our C-band flux densities and
$\alpha_{\rm C}$ demonstrate normal synchrotron emission at $\nu
\gtrsim 4$~GHz. The low L-band flux densities imply that the radio SED
becomes flatter ($\alpha$ approaches 0) near $\nu \sim 1$~GHz. This
effect was previously inferred from unusually flat (near 0) spectral
indices contrasting 1.5-to-8.4~GHz \citep{CONDON91} and directly
observed via long wavelength observations by \citet{CLEMENS10}.

For a normal star-forming galaxy, we expect the radio SED to contain
an increasing fraction of thermal emission ($\alpha = -0.1$) moving
toward higher frequencies. By $\sim 30$~GHz (the Ka band), the SED of
such a galaxy would contain roughly equal parts thermal and
non-thermal emission with flatter $\alpha$ than at $5.95$~GHz. Figure
\ref{fig:seds} shows that this simple expectation does not hold for
our data. Our Ka-band flux densities tend to lie below the simple
model: the model predicts a typical C-to-Ka ratio $\approx 2.4$, while
our median ratio is $\approx 3.15$ with $1\sigma$ scatter $0.75$. The
radio spectrum is significantly steeper than expected moving to higher
frequencies.

A steep SED at high frequencies has been previously observed in
U/LIRGs by \citet{CLEMENS08}, and the dwarf starburst NGC~1569
\citep{LISENFELD04}. If there is a significant thermal contribution at
$\nu \sim 30$~GHz, and there must be based on the high level of star
formation in these systems, then the synchrotron spectral index must
be steepening. \citet{CLEMENS08} discuss possible mechanisms for this
and \citet{CLEMENS10} revisit the topic, while \citet{LISENFELD04}
discuss possible explanations for a similar observation in the
starburst NGC 1569. Allowing that the thermal emission should be
present, the synchrotron emission must be depressed at high
frequency. Depressed synchrotron emission could result from
synchrotron and inverse Compton losses in aging starbursts or regions
with high magnetic fields. However, there is no evidence that we
observe all of these galaxies at a preferential time and these
explanations appear inadequate to match the magnitude of the
effect. Alternatively, the injection spectrum of electron energies
could differ from those in normal galaxies. This is possible for an
AGN, but less plausible for radio continuum emission originating from
star formation. We suspect that a modified injection spectrum leads to
a deficit of high frequency synchrotron emission and drives the
high-frequency flux deficit. This is a topic that we will investigate
further with our full high resolution data.

Depressed synchrotron at high frequencies may be a way to distinguish
systems where the radio is dominated by AGN. We expect a minimum level
of thermal emission in systems with radio emission due mostly to star
formation, so AGN likely provide most of the radio emission in the
steepest-spectrum sources.

\subsection{Radio-Infrared Correlation}

Despite the surprises in the radio SED, the radio and infrared fluxes
still track one another well. We calculate the far-infrared-to-radio
ratio defined by \citet{HELOU85} as $q_{\rm \nu} = \log_{10}
\left(F_{\rm FIR} / S_{\rm \nu}\right)$ with $F_{\rm FIR}$ the far-IR
flux\footnote{To calculate $q$ we use the IRAS-defined far-IR flux,
  which captures flux from 40--120$\mu$m, not the integrated IR flux
  quoted elsewhere in this paper.}  in units of $3.75 \times
10^{12}$~W~m$^{-2}$ and $S_{\nu}$ in W~m$^{-2}$~Hz$^{-1}$. The median
$q_{5.95} = 2.8$ with a scatter of $\approx 0.16$ dex. The Ka band
shows comparable scatter. The L-band $q_{1.5}$ has $\sim 50\%$ more
scatter, $\sim 0.25$ dex.  \citet{YUN01} observed 0.26~dex scatter in
the $L$-band radio-IR correlation and 0.33~dex scatter for LIRGs and
ULIRGs. The smaller overall scatter in our sample may simply be a
result of small number statistics. We interpret the lower scatter at
high frequencies in our sample as a manifestation of the same
phenomenon that leads to higher scatter in $q_{\rm 1.5}$ at high
luminosities. The curved SED at low frequencies in luminous IR
galaxies adds scatter to the radio-IR correlation as physical
processes other than just star formation become important. At higher
frequencies and in lower luminosity galaxies the effects are less
important.

The correlation among the radio bands is stronger than the correlation
with the IR, with the strongest correlation among the two synchrotron
dominated bands: $q_{\rm 5.95}$ and $q_{\rm 32.5}$ rank correlate with
$0.67$; $q_{\rm 5.95}$ and $q_{\rm 1.5}$ exhibit a rank correlation
coefficient of $0.81$; $q_{\rm 1.5}$ and $q_{\rm 32.5}$ have a rank
correlation $0.57$.

Three galaxies lie notably off the radio-IR correlation. The radio
emission from UGC~08058 (Markarian 231) is much ($4.5\sigma$) brighter
than one would derive from the IR emission. UGC~05101 shows a weaker
($2.5\sigma$) excess in the same direction. Both galaxies show
evidence for AGN in VLBI observations
\citep{LONSDALE93,LONSDALE03}. Though the contribution of the AGN to
the bolometric luminosity is debated, \citet{VEILLEUX09} estimate the
fraction of total IR emission due to the AGN to be $\sim 70\pm15\%$ in
UGC~08058 and $\sim35\pm15\%$ in UGC~05101 \citet[see][for an even
  lower estimate]{ARMUS07}.

IRASF~08572+3915 shows unusually ($2.5\sigma$) strong IR
emission. This is our faintest target and its IR-excess has been noted
before \citep[see discussion in][]{YUN04} with $\sim 25\%$ of the IR
emission in this galaxy associated with a very hot $T\gtrsim300$~K
dust component \citep[][the largest such fraction in their
  sample]{ARMUS07}. The flat spectral index at low frequencies and the
high IR-to-radio ratio in this galaxy could be naively interpreted to
indicate a very young starburst \citep[e.g.][]{ROUSSEL06}. For
example, this galaxy sits next to the youngest-age model point in the
diagrams of \citet{BRESSAN02}. However, based on near- and mid-IR
spectroscopy \citep{ARMUS07,RISALITI10} and X-Ray hardness
\citep{IWASAWA11} much of the luminosity is thought to come from a
deeply embedded AGN in the northwest nucleus. In either case, a
luminous, compact, deeply embedded object drives the luminosity and
results unusually strong IR emission. We detect it at Ka band, but not
at $>3\sigma$ over the C-band aperture so that further study will have
to wait for the the better SNR and smaller matched apertures from our
full data set.

\subsection{Tracing Embedded Star Formation} 

\begin{figure*}
\plotone{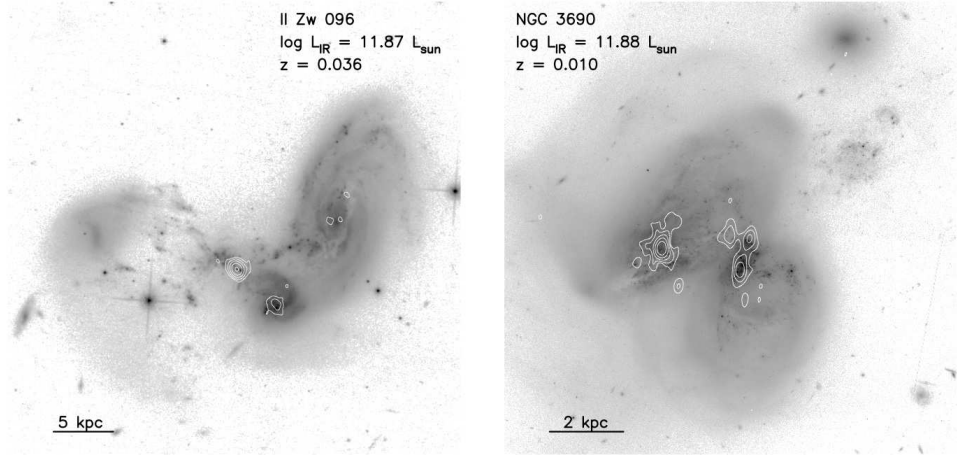}
\caption{\label{fig:overlay} EVLA Ka-band contours superimposed on
  grey scale HST ACS/WFC I-band (F814W) images of CGCG 448-020 (II Zw
  096, left) and NGC 3690 (right), both on a logarithmic stretch. The
  contours are displayed in six evenly spaced log steps between
  10--100\% of peak intensity for II Zw 096 and 2--100\% of the peak
  intensity for NGC 3690.}
\end{figure*}

Figure \ref{fig:overlay} shows our Ka-band emission superimposed on
$I$-band HST images for two targets, CGCG 448-020 (II~Zw~096, left)
and NGC~3690 (right). The images show the utility of combining
high-resolution radio data with the multiwavelength data assembled by
the Great Observatory All-sky LIRG Survey \citep[GOALS,][]{ARMUS09} to
assess the distribution and strength of embedded star formation and
AGN activity. In II Zw 096, the brightest radio continuum point does
not have an obvious optical counterpart. This bright radio emission
coincides with the peak of the IR emission measured by {\em Spitzer}
\citep{INAMI10}, underscoring that this is a star-forming peak without
an obvious optical signature. In NGC 3960 (right), several of the
radio sources have optical counterparts, but without the radio
emission it is difficult to determine which optical knots harbor
luminous embedded activity. Here too, the 32.5~GHz continuum peaks
coincide with the bright IR emission \citep{ALONSO-HERRERO10} but with
higher angular resolution.

\subsection{Summary} 

We present new measurements of radio continuum emission from 22
U/LIRGs across 1~GHz windows centered at 4.7, 7.2, 29, and 36~GHz.
Our observations are consistent with typical synchrotron emission at
4.2--7.7~GHz. Comparing these normal C-band results with new
high-frequency 29--36~GHz fluxes and literature measurements at L, X,
and K-bands, we see that the radio SEDs of U/LIRGs systematically
deviate from a simple two-component model at both low and high
frequencies. The new EVLA data here provide: 1) observations filling
in previously unsampled parts of the radio SED; 2) a clear observation
of ``normal'' synchrotron behavior in the range 4--8~GHz, reinforcing
that the radio SED must be curving away from simple expectations; and
3) an observation of systematically low high-frequency flux density at
higher frequencies than have been previously studied. The radio fluxes
correlate with IR emission as expected, with some suggestion that the
C- and Ka-bands exhibit a more direct link to IR emission than the
L-band. 

An ongoing extension of this project to all EVLA configurations will
dramatically improve our resolution and signal-to-noise ratio in both
frequency bands. Combining these data with the multiwavelength data
available for our targets, we expect to test if the star formation
surface density exhibits any maximum value \citep[e.g., the limit set
  by momentum-driven winds suggested by][]{MURRAY05}, derive
characteristic sizes for the star-forming disks, map the spectral
index at $\lesssim 0.5\arcsec$ resolution, and explore the
implications of the brightness temperature and radio SED for the
dominant radio emission mechanism.

\acknowledgments We thank the anonymous referee for suggestions that
improved this work. The National Radio Astronomy Observatory is a
facility of the National Science Foundation operated under cooperative
agreement by Associated Universities, Inc. Support for AKL during part of this work was
provided by NASA through Hubble Fellowship grant HST-HF-51258.01-A
awarded by the Space Telescope Science Institute, which is operated by
the Association of Universities for Research in Astronomy, Inc., for NASA,
under contract NAS 5-26555.

\begin{deluxetable*}{llcccccccccc}
\tabletypesize{\footnotesize}
\tablecaption{\label{tab:results}
Integrated Properties} 
\tablehead{
\colhead{Galaxy} & 
\colhead{Dist.\tablenotemark{a}} &
\colhead{$L_{\rm IR}$\tablenotemark{a}} & 
\colhead{$S_{5.95}$\tablenotemark{b}} &
\colhead{$\alpha_{C}$\tablenotemark{c}} &
\colhead{$S_{32.5}$\tablenotemark{b}} &
\colhead{$S_{1.49}$ (C91)\tablenotemark{d}} &
\colhead{$S_{8.44}$ (C91)\tablenotemark{d}} &
\colhead{$S_{15}$ (C08)\tablenotemark{d}} \\
\colhead{} & 
\colhead{[Mpc]} & 
\colhead{[$\log_{10} L_\odot$]} &
\colhead{[mJy]} &
\colhead{} &
\colhead{[mJy]} &
\colhead{[mJy]} &
\colhead{[mJy]} &
\colhead{[mJy]}
}
\startdata
CGCG 436-030 & 134 & 11.69 & $18.6 \pm 0.04$ & $-0.72$ & $8.6 \pm 0.6$ & 43.1& 12.7 & 8.73 \\
CGCG 448-020 (II~Zw~096) & 161 &  11.94 & $14.6 \pm 0.06$ & $-0.66$ & $6.8 \pm 1.1$ & \nodata & \nodata & \nodata \\
III Zw 035 & 119 & 11.64 & $25.4 \pm 0.05$ & $-0.50$ & $7.3 \pm 0.6$ & 39.3 & 19.7 & 9.7 \\
IRAS 19542+1110 & 295 & 12.12 & $9.5 \pm 0.04$ & $-0.79$ & $3.4 \pm 0.7$ & \nodata & \nodata & \nodata \\
IRAS 21101+5810 & 174 & 11.81 & $9.8 \pm 0.04$ & $-0.67$ & $4.6 \pm 0.6$ & \nodata & \nodata & \nodata \\
\\
IRAS F01365-1042 & 210 & 11.85 & $10.0 \pm 0.04$ & $-0.47$ & $4.1 \pm 0.5$ & 17.0 & 8.2 & 3.97 \\
IRAS 0857+39 & 264 & 12.16 & $4.44 \pm 0.04$ & $-0.29$ & \nodata\tablenotemark{e} & 6.5 & 4.1 & 3.18 \\
IRAS 1525+36 & 254 & 12.08 & $12.0 \pm 0.04$ & $-0.44$ & $4.3 \pm 0.8$ & 12.8 & 10.5 & 4.2 \\
IRAS 1713+53 & 232 & 11.96 & $9.3 \pm 0.04$ & $-0.78$ & $4.1 \pm 0.7$ & 28.4 & 8.9 & \nodata \\
IRAS F23365+3604 & 287 & 12.20 & $10.6 \pm 0.04$ & $-0.72$ & $2.4 \pm 0.5$ & \nodata & \nodata & \nodata\\
\\
MCG +07-23-019 (Arp 148) & 158 & 11.62 & $16.0 \pm 0.06$ & $-0.66$ & $5.1 \pm 1.2$ & \nodata & \nodata & \nodata\\
NGC 3690 (Mrk 171) & 51 & 11.93 & $275.5 \pm 0.34$ & $-0.62$ & $89.0 \pm 2.6$ & 658.0 & \nodata & \nodata \\
UGC 04881 (Arp 55) & 178 & 11.74 & $11.4 \pm 0.09$ & $-0.96$ & $3.1 \pm 1.0$ & 29.0 & 8.8 & 3.5 \\
UGC 05101 & 177 & 12.01 & $61.5 \pm 0.08$ & $-0.75$ & $16.8 \pm 1.1$ & 146.0 & 52.6 & 18.0 \\
UGC 08058 (Mrk 231) & 192 & 12.57 & $312.8 \pm 0.2$ & $-0.17$ & $88 \pm 3.5$ & 240.0 & 265.0 & \nodata \\
\\
UGC 08387 (Arp 193, I Zw 056) & 110 & 11.73 & $46.3 \pm 0.08$ & $-0.55$ & $16.5 \pm 0.9$ & 106.0 & 34.9 & 7.5 \\
UGC 08696 (Mrk 273, I Zw 071) & 173 & 12.21 & $60.3 \pm 0.08$ & $-0.72$ & $19.5 \pm 1.0$ & 130.0 & 43.5 & 14.3 \\
UGC 09913 (Arp 220) & 88 & 12.28 & $194.5 \pm 0.08$ & $-0.59$ & $59.5 \pm 2.5$ & 301.1 & 148.0 & \nodata \\
VII Zw 031 & 240 & 11.99 & $12.5 \pm 0.04$ & $-0.91$ & $3.7 \pm 0.7$ & \nodata & \nodata & \nodata \\
VV 250a & 142 & 11.81 & $19.6 \pm 0.05$ & $-0.69$ & $3.8 \pm 1.2$ & 51.2 & 17.0 & \nodata \\
\\
VV 340a & 157 & 11.74 & $23.7 \pm 0.12$ & $-1.06$ & \nodata\tablenotemark{e} & 68.8 & \nodata & \nodata \\
VV 705 & 183 & 11.92 & $19.6 \pm 0.05$ & $-0.75$ & $4.8 \pm 0.8$ & 46.8 & 12.1 & 3.6
\enddata
\tablenotetext{a}{Hubble flow distance and IR luminosity adopted from \citet{SANDERS03} following \citet{ARMUS09}.}
\tablenotetext{b}{Flux density integrated over significant C-band emission. See text. Uncertainties are {\em only} statistical, bootstrapped from the map.}
\tablenotetext{c}{Intensity-weighted in-band spectral index derived from MFS imaging. See text.}
\tablenotetext{d}{Literature flux densities from \citet{CONDON91} and \citet{CLEMENS08}.}
\tablenotetext{e}{Signal-to-noise ratio $<3$ integrated over the C-band beam. Still detect at Ka resolution.}
\end{deluxetable*}

\acknowledgments

\end{document}